\documentclass[journal]{IEEEtran}

\usepackage{lipsum}

\usepackage{cite}

\ifCLASSINFOpdf
   \usepackage[pdftex]{graphicx}
\else
\fi
\usepackage{multirow}

\usepackage{array}

\ifCLASSOPTIONcompsoc
 \usepackage[caption=false,font=normalsize,labelfont=sf,textfont=sf]{subfig}
\else
  \usepackage[caption=false,font=footnotesize]{subfig}
\fi
\usepackage{url}

\usepackage{amsmath}
\usepackage{amssymb}
\usepackage{amsfonts}
\usepackage{esint}
\usepackage{mathrsfs}
\usepackage{amsthm}
\usepackage{amscd}
\usepackage{longtable}
\usepackage{listings}
\usepackage{booktabs,longtable}
\usepackage{graphicx}
\usepackage{multirow}
\usepackage{tikz}

\newcommand{\vect}[1]{\mathbf{#1}}

\newcommand{\matr}[1]{\mathbf{#1}}

\newcommand{\junk}[1] {}

\def\XXint#1#2#3{{\setbox0=\hbox{$#1{#2#3}{\int}$}
\vcenter{\hbox{$#2#3$}}\kern-.5\wd0}}

\newcommand*\widebar[1]{%
  \hbox{%
    \vbox{%
      \hrule height 0.5pt 
      \kern0.3ex
      \hbox{%
        \kern-0.05em
        \ensuremath{#1}%
        \kern-0.05em
      }%
    }%
  }%
} 
\begin{document}

%
\title{TurboMOR: an Efficient Model Order Reduction Technique for
RC Networks with Many Ports}
%
%
%

\author{Denis~Oyaro,~\IEEEmembership{Student~Member,~IEEE,}
        and~Piero~Triverio,~\IEEEmembership{Member,~IEEE}\\~\\
        Submitted for publication on the IEEE Transactions on Computer-Aided Design of Integrated Circuits and Systems on July 1st, 2015        
\thanks{This work was supported in part by the Natural Sciences and Engineering Research Council of
Canada (Discovery Grant program) and in part by the Canada Research Chairs program.}
\thanks{D. Oyaro and P.~Triverio are with the Edward S. Rogers Sr. Department of Electrical and Computer Engineering, University of Toronto, Toronto, M5S 3G4 Canada (email: piero.triverio@utoronto.ca).}
}

\markboth{Oyaro and Triverio - TurboMOR}%
{Oyaro and Triverio - TurboMOR}%



\maketitle

\begin{abstract}
Model order reduction (MOR) techniques play a crucial role in the computer-aided design of modern integrated circuits, where they are used to reduce the size of parasitic networks. Unfortunately, the efficient reduction of passive networks with many ports is still an open problem. Existing techniques do not scale well with the number of ports, and lead to dense reduced models that burden subsequent simulations. In this paper, we propose TurboMOR, a novel MOR technique for the efficient reduction of passive RC networks. TurboMOR is based on moment-matching, achieved through efficient congruence transformations based on Householder reflections. A novel feature of TurboMOR is the block-diagonal structure of the reduced models, that makes them more efficient than the dense models produced by existing techniques. Moreover, the model structure allows for an insightful interpretation of the reduction process in terms of system theory. Numerical results show that TurboMOR scales more favourably than existing techniques in terms of reduction time, simulation time and memory consumption.
\end{abstract}

\begin{IEEEkeywords}
Model order reduction, many ports, moment matching, parasitics, partitioning.
\end{IEEEkeywords}

\IEEEpeerreviewmaketitle

\section{Introduction}

\IEEEPARstart{W}{hile} designing VLSI chips, engineers need to take into account the parasitic resistance, capacitance and inductance of signal- and power-delivery interconnects, in order to prevent signal and power integrity issues~\cite{chen1998interconnect,silva2007issues,nassif2008power}. 
Electromagnetic solvers are used to extract RC or RLC interconnect models, which are then connected to  non-linear devices for system-level simulations. Unfortunately, parasitic networks  can be very large, featuring a huge number of components, nodes and ports. Direct simulation involving such large networks is often prohibitive. Model order reduction (MOR) is frequently used to reduce parasitic models to a manageable size, and accelerate subsequent simulations.

Several approaches to MOR have been proposed in the last decades, such as node elimination~\cite{sheehan1999ticer}, Krylov subspaces~\cite{grimme1997krylov,celik2002ic}, and balancing~\cite{antoulas2005approximation}. Krylov methods are widely used for parasitic reduction, since they are more scalable than balancing methods. 
Among them, PRIMA~\cite{odabasioglu1997prima} is one of the most popular and widely used Krylov algorithms. PRIMA's success is due to its ability to guarantee the passivity of the ROM, a mandatory property to prevent divergent transient simulations~\cite{triverio2007stability}. Unfortunately, PRIMA can become very inefficient when applied to networks with many ports.
PRIMA generates the reduced model through a congruence transformation with an orthogonal matrix that spans a suitable Krylov subspace. The orthogonal projection matrix is dense and can become very large when ports are many. Generating the ROM becomes very time consuming, since it involves products between large and dense matrices. In some cases, even storing the projection matrix can be challenging. Moreover, the obtained reduced model is dense, large and frequently slower than the original system. These issues affect most existing techniques and are an outstanding issue in MOR~\cite{silva2007outstanding}.

A number of techniques have been recently proposed to address such challenges. Methods like SVDMOR~\cite{feldmann2004model}, ESVDMOR~\cite{liu2008efficient}, RECMOR~\cite{feldmann2004sparse} and several others~\cite{benner2010model,li2006model} aim at reducing the number of ports before applying PRIMA. This is done by exploiting the correlation that may exist between different ports. However, practical networks with many ports rarely exhibit a high degree of correlation~\cite{yan2012decentralized}. 

In~\cite{benner2008using,zhang2011block,nouri2013efficient}, the problem of reducing networks with many ports is simplified by clustering inputs into small groups, and reducing each subsystem individually. These methods generate accurate and block diagonal ROMs that are sparse. However, since subsystems are treated independently, passivity is not always guaranteed.
 
Another method known as SIP~\cite{ye2008sparse} offers a more efficient approach to moment matching for RC networks. Rather than explicitly constructing the projection matrix, sparse matrix manipulations are used to generate the reduced matrices directly using the Schur complement, an idea also used in PACT~\cite{kerns1997stable}. This makes SIP more efficient than PRIMA for large networks with many ports. However, SIP can match only two moments per expansion point. This level of accuracy is not always sufficient for practical applications~\cite{ye2008sparse}, as we will show in Sec.~\ref{NumericalResults}. The authors in~\cite{ye2008sparse} suggest using multi-point moment matching~\cite{grimme1997krylov,celik2002ic,tan2007advanced} to achieve more accuracy. However, the obtained reduced matrices can be singular, and avoiding this issue does not seem to be trivial.

In~\cite{ionuctiu2011sparserc}, the SparseRC method is proposed, combining graph-partitioning techniques~\cite{miettinen2011partmor} with a SIP-like reduction process. A divide and conquer strategy is used to partition the original system into smaller subsystems, then reduced separately with a method similar to SIP~\cite{ye2008sparse}. The resulting ROM has the same partitioned structure as the original system. Such a reduction strategy is efficient in terms of memory and cpu time for large networks, since the problem of reducing the large system simplifies to reducing smaller subsystems that can be managed efficiently. The generated ROM is also sparse. SparseRC, however, like SIP, is limited to matching two moments per expansion point. While PRIMA can be used to match additional moments, as suggested in~\cite{ionuctiu2011sparserc}, this reduces efficiency, because of the limitations of PRIMA discussed previously.

In this paper, we propose TurboMOR, a novel MOR technique for RC networks with many ports. TurboMOR achieves moment-matching without explicitly computing a dense projection matrix as in PRIMA. Efficient and memory-conscious Householder reflections~\cite{golub2012matrix} are used to generate the reduced model, and match two moments per iteration. Differently from previous methods such as SIP~\cite{ye2008sparse}, an arbitrary number of moments can be matched, providing full control on accuracy. TurboMOR can be combined with partitioning~\cite{ionuctiu2011sparserc,miettinen2011partmor} to reduce very large networks. A key feature of TurboMOR is the block-diagonal structure of the reduced models, that addresses the poor efficiency of the dense models produced by existing moment-matching techniques. The block diagonal structure also lends itself to a novel and insightful interpretation of moment matching in terms of cascaded subsystems. The reduced models produced by TurboMOR are passive, retain the input-output structure of the original system, and can be synthesized into an equivalent RC netlist~\cite{yang2007rlcsyn}. Numerical tests demonstrates the superior scalability of TurboMOR in terms of reduction time, simulation time, and memory consumption.

The rest of the paper is organized as follows. In Sec.~\ref{ProblemFormulation}, we state the problem and briefly review the foundations of moment matching. In Sec.~\ref{TurboMOR}, we discuss the theoretical derivation and practical implementation of TurboMOR. Sec.~\ref{NumericalResults} compares TurboMOR against the state of the art. In Sec.~\ref{Conclusion} we draw our conclusions, and in the Appendix we provide some mathematical proofs.

\section{Problem Formulation}\label{ProblemFormulation}

We consider a passive network made by resistors and capacitors with $m$ nodes and $p$ ports. Using nodal analysis~\cite{ho1975modified}, the network can be described in the Laplace domain by the systems of equations
\begin{equation}\label{eq:solve1}
\left \{
\begin{aligned}
\matr{G}\vect{x}(s) + s\matr{C}\vect{x}(s) &= \matr{B}\vect{u}(s)\\
\vect{y}(s)&=\matr{B}^{T}\vect{x}(s)
\end{aligned}
\right.
\end{equation}
where vectors $\vect{u}(s) \in \mathbb{R}^p$ and $\vect{y}(s) \in \mathbb{R}^p$ collect all port currents and port voltages, respectively. Vector $\vect{x}(s)  \in \mathbb{R}^m$ contains all nodal voltages.
Matrices $\matr{G}$, $\matr{C} \in \mathbb{R}^{m \times m}$ are conductance and capacitance matrices, respectively. They are symmetric and non-negative definite. Matrix $\matr{B} \in \mathbb{R}^{m \times p}$ maps input ports to the nodal equations, and $^T$ denotes transposition. The transfer function of \eqref{eq:solve1} reads
\begin{equation}\label{eq:solve2}
\matr{H}(s)=\matr{B}^T(\matr{G}+s\matr{C})^{-1}\matr{B}
\end{equation}
The goal of MOR is to approximate~\eqref{eq:solve1} with a model of much lower order $n \ll m$
\begin{equation}\label{eq:solve3}
\left \{
\begin{aligned}
\hat{\matr{G}}\hat{\vect{x}}(s) + s\hat{\matr{C}}\hat{\vect{x}}(s)&=\hat{\matr{B}}\vect{u}(s)\\
\hat{\vect{y}}(s)&=\hat{\matr{B}}^{T} \hat{\vect{x}}(s) 
\end{aligned}
\right.
\end{equation}
where $\hat{\matr{G}}$, $\hat{\matr{C}} \in \mathbb{R}^{n \times n}$, $\hat{\matr{B}} \in \mathbb{R}^{n \times p}$ and $\hat{\vect{x}}(s) \in \mathbb{R}^n$. This model must accurately capture the response of the original system across the frequency range of interest. 

One way of ensuring accuracy is through Pad\'e approximation, also known as moment matching. Around $s=0$, the Taylor series expansion of~\eqref{eq:solve2} reads
\begin{equation}
\matr{H}(s)=\matr{M}_0+\matr{M}_1s+\matr{M}_2s^2+\dots
\label{eq:M}
\end{equation}
The coefficients $\matr{M}_k$ are called moments of~\eqref{eq:solve1} at DC~\cite{grimme1997krylov,celik2002ic,odabasioglu1997prima}, and can be related to the systems matrices as
\begin{equation}\label{eq:solve5}
\matr{M}_k=\matr{B}^T(-\matr{G}^{-1}\matr{C})^k\matr{G}^{-1}\matr{B} \quad \forall k=0,1,2,...
\end{equation}
The moments of the reduced model are defined similarly, as the Taylor expansion coefficients of the transfer function
\begin{equation}\label{eq:solve4}
\hat{\matr{H}}(s)=\hat{\matr{B}}^T(\hat{\matr{G}}+s\hat{\matr{C}})^{-1}\hat{\matr{B}}
\end{equation}
of reduced model~\eqref{eq:solve3}.

The goal of moment matching is to generate a ROM~\eqref{eq:solve3}
that will match the first moments of the original system
\begin{equation}
\matr{M}_k=\hat{\matr{M}}_k \quad \forall k=0,...\,,2q-1
\end{equation} 
up to a given order controlled by $q$. Since, for RC networks, moments are typically matched in pairs, we denote the number of  matched moments as $2q$. By increasing $q$ the ROM will become more accurate, but also larger.

In PRIMA, moment matching is performed with a congruence transformation applied to the matrices of the original system~\eqref{eq:solve1}
\begin{equation}\label{eq:prima1}
\hat{\matr{G}}=\matr{Q}^T\matr{G}\matr{Q}, \hspace{3mm} \hat{\matr{C}}=\matr{Q}^T\matr{C}\matr{Q}, \hspace{3mm} \hat{\matr{B}}=\matr{Q}^T\matr{B}
\end{equation}
The columns of $\matr{Q} \in \mathbb{R}^{m \times qp}$ span the Krylov subspace
\begin{equation}
\mathcal{K}_q(\matr{A},\matr{R})=span\{\matr{R},\matr{A}\matr{R},\matr{A}^2\matr{R},...,\\\matr{A}^{2q-1}\matr{R}\}
\end{equation}
where $\matr{A}=-\matr{G}^{-1}\matr{C}$ and $\matr{R}=\matr{G}^{-1}\matr{B}$. It can be shown that ROM~\eqref{eq:prima1} matches the first $2q$ moments of the original system. The reduced model is of size $n=qp$, and is passive by construction since congruence transformation~\eqref{eq:prima1} maintains the non-negative nature of $\matr{G}$ and $\matr{C}$. The projection matrix $\matr{Q}$ is constructed numerically with the block Arnoldi process~\cite{celik2002ic}, an orthogonalization procedure similar to the modified Gram–Schmidt process~\cite{golub2012matrix}. Unfortunately, orthogonalization leads to a dense $\matr{Q}$. As a result, when $p$ is high, computing $\matr{Q}$ and projection products~\eqref{eq:prima1} can be very expensive. For very large networks, even storing $\matr{Q}$ becomes an issue, since its size can easily exceed several Gigabytes. Moreover, transformations~\eqref{eq:prima1} lead to a dense ROM, which will burden any subsequent circuit simulation. These bottlenecks, which make existing methods quite inefficient for many-port networks, are tackled by the proposed method.

\section{Proposed Method} 
\label{TurboMOR}

In this section, we discuss the theoretical derivation of TurboMOR and how it can be implemented for maximum efficiency. The method works recursively, matching two moments per iteration. We discuss the first two iterations in detail, before generalizing. 

\subsection{Theoretical Derivation}
\label{sec:theory}

\subsubsection{Matching Two Moments}
The first iteration of the proposed method is analogous to~\cite{kerns1997stable,ye2008sparse,ionuctiu2011sparserc}. Nodes are first reordered in such a way that port nodes come first, followed by internal nodes. After reordering, system~\eqref{eq:solve1} reads
\begin{subequations}
\begin{gather}
\left(
\begin{bmatrix} 
\matr{G}_{11} & \ast \\
\matr{G}_{21} & \matr{G}_{22}
\end{bmatrix}
+ s\begin{bmatrix}
\matr{C}_{11} & \ast \\
\matr{C}_{21} &\matr{C}_{22}
\end{bmatrix} \right)
\begin{bmatrix}
\vect{x}_1 \\
\vect{x}_2
\end{bmatrix}=\begin{bmatrix}
\matr{B}_{1} \\
\matr{0}
\end{bmatrix}\vect{u}
\label{eq:solve7a}\\
\vect{y}= \begin{bmatrix}
\matr{B}_{1}^T & \matr{0}
\end{bmatrix}\begin{bmatrix}
\vect{x}_1 \\
\vect{x}_2
\end{bmatrix} 
\label{eq:solve7b}
\end{gather}
\end{subequations}
where $\vect{x}_1 \in \mathbb{R}^p$ and $\vect{x}_2 \in \mathbb{R}^{m-p}$ denote port and internal node voltages, respectively. The symbol $\ast$ is used in symmetric matrices to denote the transpose of the symmetric block across the diagonal. For the purpose of shortening our notation, we do not indicate explicitly the dependency on $s$ for input, output and state variables. Submatrix $\matr{G}_{21}$ describes the resistive couplings present between internal and port nodes. We eliminate this block through Gaussian elimination, using the congruence transformation~\eqref{eq:prima1} with $\matr{Q}$ given by
\begin{equation}
\matr{Q}^{(1)} = \begin{bmatrix}
\matr{I}_p & \matr{0} \\ 
-\matr{K}^{-T}\matr{K}^{-1}\matr{G}_{21} & \matr{I}_{m-p}
\end{bmatrix}
\label{eq:congruence}
\end{equation}
Matrix $\matr{K}$ is the Cholesky factor~\cite{golub2012matrix} of $\matr{G}_{22}$. For the time being, we assume $\matr{G}_{22}$ to be positive definite (strictly). In Sec.~\ref{sec:singularG}, we will discuss how a singular $\matr{G}_{22}$ can be handled. Matrix $\matr{I}_p$ is the identity matrix of size $p \times p$. After the congruence, equations~\eqref{eq:solve7a} and~\eqref{eq:solve7b} become
\begin{subequations}
\begin{gather}
\left(
\begin{bmatrix}
\matr{G}_{11}^{(1)} & \matr{0} \\
\matr{0} & \matr{G}_{22}
\end{bmatrix}
+ s\begin{bmatrix}
\matr{C}_{11}^{(1)} & \ast \\
\matr{C}_{21}^{(1)} & \matr{C}_{22}
\end{bmatrix} \right)
\begin{bmatrix}
\vect{x}_1 \\
\vect{x}_2^{(1)}
\end{bmatrix}=\begin{bmatrix}
\matr{B}_1 \\
\matr{0}
\end{bmatrix}\vect{u}
\label{eq:solve8a}
\\
\vect{y}=\begin{bmatrix}
\matr{B}_{1}^T & \matr{0}
\end{bmatrix}\begin{bmatrix}
\vect{x}_1 \\
\vect{x}_2^{(1)}
\end{bmatrix}
\label{eq:solve8b}
\end{gather}
\end{subequations}
where
\begin{align}
\matr{G}_{11}^{(1)}&=\matr{G}_{11} - \matr{G}_{21}^T\matr{K}^{-T}\matr{K}^{-1}\matr{G}_{21} \label{eq:solve10c} \\
\matr{C}_{11}^{(1)}&=\matr{C}_{11} - \matr{G}_{21}^T\matr{K}^{-T}\matr{K}^{-1}\matr{C}_{21} - \matr{C}_{21}^T\matr{K}^{-T}\matr{K}^{-1} \matr{G}_{21} \notag \\ &\hspace{20mm} + \matr{G}_{21}^T\matr{K}^{-T}\matr{K}^{-1}\matr{C}_{22}\matr{K}^{-T}\matr{K}^{-1}\matr{G}_{21} \label{eq:solve10a} \\ 
\matr{C}_{21}^{(1)}&=\matr{C}_{21} - \matr{C}_{22}\matr{K}^{-T}\matr{K}^{-1}\matr{G}_{21} \label{eq:solve10b} 
\end{align}
With Gaussian elimination, all resistive couplings between port nodes and internal nodes have been eliminated, leaving only capacitive couplings.

\begin{figure}[!t]
\centering
\begin{tikzpicture}[>=latex]\sffamily
        \node[draw,fill=white!20!white,minimum height=1cm,minimum width=2cm] (sigOne) at(0,0){$\Sigma_{1}^{(1)}$};
        \node[draw,fill=white!20!white,minimum height=1cm,minimum width=2cm] (sigTwo) at(0,-2.5){$\Sigma_{2}^{(1)}$};
				 \node[draw,fill=white!20!white,minimum height=0.8cm,minimum width=0.6cm] (sOne) at(-2.5,-1.25){$\frac{d}{dt}$};
				\node[draw,fill=white!20!white,minimum height=0.8cm,minimum width=0.6cm] (sTwo) at(2.3,-1.25){$\frac{d}{dt}$};
        \draw[->] (sOne.north)|-node[below,pos=0.8]{$\matr{u}_{1}^{(1)}$}([yshift=-3mm]sigOne.west);
				\draw[->] (sigTwo.west)-|node[above,pos=0.2]{$\matr{y}_{2}^{(1)}$}(sOne.south);
				\draw[->] (sTwo.south)|-node[above,pos=0.8]{$\matr{u}_{2}^{(1)}$}(sigTwo.east);
				\draw[->] (sigOne.east)-|node[above,pos=0.2]{$\vect{x}_{1}$}coordinate(temp)(sTwo.north);
				\draw[<-]([yshift=3mm]sigOne.west)--node[above,pos=0.7]{$\matr{u}$}+(-2cm,0); 
				\draw[->](temp)--+(1.3,0)node[right]{$\matr{y}$};
\end{tikzpicture}
\caption{System theory interpretation of~\eqref{eq:solve7a} and~\eqref{eq:solve7b}. The original system has been decomposed into two subsystems $\Sigma_{1}^{(1)}$ and $\Sigma_{2}^{(1)}$, decoupled at DC.}
\label{fig:interaction1}
\end{figure}
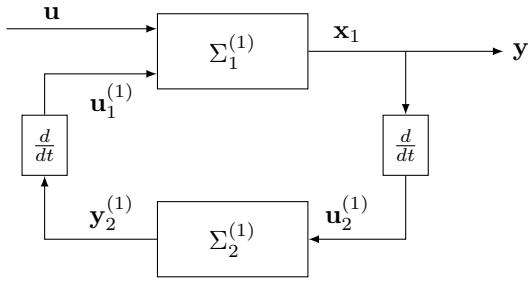
The obtained equations lend themselves to a useful interpretation in terms of system theory, depicted in Fig.~\ref{fig:interaction1}. System~\eqref{eq:solve8a}-\eqref{eq:solve8b}  can be seen as the cascade of  a system $\Sigma_{1}^{(1)}$ of order $p$
\begin{equation} \label{eq:solve11a}
\Sigma_{1}^{(1)}:\left \{
\begin{aligned}
\matr{G}_{11}^{(1)}\vect{x}_{1} + s\matr{C}_{11}^{(1)}\vect{x}_{1}&=\vect{u}_{1}^{(1)} + \matr{B}_1\vect{u}\\
\vect{y}&=\matr{B}_{1}^T \vect{x}_{1}
\end{aligned}
\right.
\end{equation}
and a system $\Sigma_{2}^{(1)}$ of order $m-p$
\begin{equation}\label{eq:solve11}
\Sigma_{2}^{(1)}:\left\{
\begin{aligned}
\matr{G}_{22}\vect{x}_{2}^{(1)} + s\matr{C}_{22}\vect{x}_{2}^{(1)}&=-\matr{C}_{21}^{(1)}\vect{u}_{2}^{(1)}\\
\vect{y}_{2}^{(1)}&=-(\matr{C}_{21}^{(1)})^T\vect{x}_{2}^{(1)}
\end{aligned}
\right.
\end{equation}
Only the first subsystem $\Sigma_{1}^{(1)}$ is directly connected to the input/output ports of the network. Subsystem $\Sigma_{2}^{(1)}$ is instead connected only to $\Sigma_{1}^{(1)}$, through equations $\vect{u}_{1}^{(1)}= s\vect{y}_{2}^{(1)}$ and $\vect{u}_{2}^{(1)}=s\vect{x}_{1}$, which define time derivatives. The coupling between the two  subsystems is thus \emph{purely dynamical}. At DC, the second system is completely decoupled from $\Sigma_{1}^{(1)}$ and the network ports, and has no influence on the transfer function $\matr{H}(s)$ between input $\vect{u}$ and output $\vect{y}$. At low frequency, the coupling between the two is weak, and the overall system response is  given  mainly by $\Sigma_{1}^{(1)}$. Therefore, the first subsystem alone can be interpreted as a ROM of order $p$ of the original system
\begin{equation}\label{eq:solve12}
\left\{
\begin{aligned}
\matr{G}_{11}^{(1)}\vect{x}_{1} + s\matr{C}_{11}^{(1)}\vect{x}_{1}&=\matr{B}_1\vect{u}\\
\hat{\vect{y}}&=\matr{B}_{1}^T \vect{x}_{1}
\end{aligned} 
\right.
\end{equation} 
In the Appendix, we indeed prove that~\eqref{eq:solve12} matches  the first two moments of the original system at $s=0$. From an accuracy standpoint, the proposed ROM is thus equivalent in size and accuracy to the ROMs generated by other moment matching techniques. Its computation, however, requires less effort, since its matrices~\eqref{eq:solve10c} and~\eqref{eq:solve10a} can be computed cheaply using sparse matrix techniques.

\subsubsection{Matching Four Moments}

In order to match more than two moments, the presence of $\Sigma_{2}^{(1)}$ must be taken into account. Instead of applying PRIMA to $\Sigma_{2}^{(1)}$ as in~\cite{ionuctiu2011sparserc}, loosing efficiency, we show how  additional moments can be efficiently matched by further decomposing $\Sigma_{2}^{(1)}$. 

First, we apply a congruence transformation to~\eqref{eq:solve11} using $\matr{Q} = \matr{K}^{-T}$ in~\eqref{eq:prima1}
\begin{equation}\label{eq:solve13}
\left\{
\begin{aligned}
\matr{I}_{m-p}\vect{z}_{2}^{(1)} + s\matr{K}^{-1}\matr{C}_{22}\matr{K}^{-T}\vect{z}_{2}^{(1)}&=-\matr{K}^{-1}\matr{C}_{21}^{(1)}\vect{u}_{2}^{(1)}\\
\vect{y}_{2}^{(1)}&=-(\matr{C}_{21}^{(1)})^T\matr{K}^{-T}\vect{z}_{2}^{(1)}
\end{aligned}
\right.
\end{equation}
where $\vect{x}_{2}^{(1)}=\matr{K}^{-T}\vect{z}_{2}^{(1)}$. This step turns $\matr{G}_{22}$ into the identity matrix, and does not require expensive computations since $\matr{K}$ is already available from the previous iteration.

Then, with a series of Householder reflections~\cite{golub2012matrix},  we compute the QR factorization of the input-to-state matrix in~\eqref{eq:solve13}
\begin{equation}
(\matr{Q}^{(2)})^T\matr{K}^{-1}\matr{C}_{21}^{(1)}= \begin{bmatrix}
\matr{R}^{(2)}\\
\matr{0}
\end{bmatrix}
\label{eq:qr}
\end{equation}
where $\matr{R}^{(2)} \in \mathbb{R}^{p \times p}$ is upper triangular and  $\matr{Q}^{(2)} \in \mathbb{R}^{(m-p) \times (m-p)}$ is an orthogonal matrix given by the product of Householder reflectors~\cite{golub2012matrix}. 

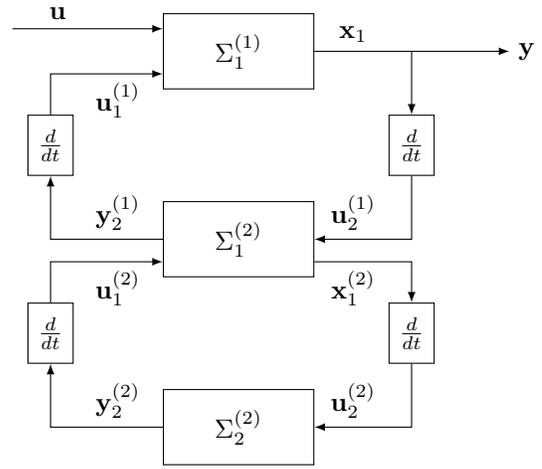
\begin{figure}[!t]
\centering
\begin{tikzpicture}[>=latex]\sffamily
        \node[draw,fill=white!20!white,minimum height=1cm,minimum width=2cm] (sigOne) at(0,0){$\Sigma_{1}^{(1)}$};
        \node[draw,fill=white!20!white,minimum height=1cm,minimum width=2cm] (sigTwo) at(0,-2.5){$\Sigma_{1}^{(2)}$};
		\node[draw,fill=white!20!white,minimum height=1cm,minimum width=2cm] (sigThree) at(0,-5){$\Sigma_{2}^{(2)}$};
				 \node[draw,fill=white!20!white,minimum height=0.8cm,minimum width=0.6cm] (sOne) at(-2.5,-1.25){$\frac{d}{dt}$};
				\node[draw,fill=white!20!white,minimum height=0.8cm,minimum width=0.6cm] (sTwo) at(2.3,-1.25){$\frac{d}{dt}$};
				 \node[draw,fill=white!20!white,minimum height=0.8cm,minimum width=0.6cm] (sThree) at(-2.5,-3.75){$\frac{d}{dt}$};
				\node[draw,fill=white!20!white,minimum height=0.8cm,minimum width=0.6cm] (sFour) at(2.3,-3.75){$\frac{d}{dt}$};				
        \draw[->] (sOne.north)|-node[below,pos=0.8]{$\matr{u}_{1}^{(1)}$}([yshift=-3mm]sigOne.west);
				\draw[->] (sigTwo.west)-|node[above,pos=0.2]{$\matr{y}_{2}^{(1)}$}(sOne.south);
				\draw[->] (sTwo.south)|-node[above,pos=0.8]{$\matr{u}_{2}^{(1)}$}(sigTwo.east);
				\draw[->] (sigOne.east)-|node[above,pos=0.2]{$\vect{x}_{1}$}coordinate(temp)(sTwo.north);
				\draw[<-]([yshift=3mm]sigOne.west)--node[above,pos=0.7]{$\matr{u}$}+(-2cm,0); 
				\draw[->](temp)--+(1.3,0)node[right]{$\matr{y}$};    
				\draw[->] ([yshift=-3mm] sigTwo.east)-|node[below,pos=0.2]{$\vect{x}_{1}^{(2)}$}coordinate(temp)( sFour.north);
        \draw[->] (sThree.north)|-node[below,pos=0.8]{$\matr{u}_{1}^{(2)}$}([yshift=-3mm]sigTwo.west);
				\draw[->] (sigThree.west)-|node[above,pos=0.2]{$\matr{y}_{2}^{(2)}$}(sThree.south);
				\draw[->] (sFour.south)|-node[above,pos=0.8]{$\matr{u}_{2}^{(2)}$}(sigThree.east);				
\end{tikzpicture}
\caption{Structure of the system obtained after two iterations of the proposed method.}
\label{fig:interaction2}
\end{figure}

After $\matr{Q}^{(2)}$ is applied to~\eqref{eq:solve13} with a congruence transformation, the system will read
\begin{subequations}
\begin{gather}
\left(
\begin{bmatrix}
\matr{I}_{p} & \matr{0} \\
\matr{0} & \matr{I}_{m-2p}
\end{bmatrix}
+ s\begin{bmatrix}
\matr{C}_{11}^{(2)} & \ast \\
\matr{C}_{21}^{(2)} & \matr{C}_{22}^{(2)}
\end{bmatrix} \right)
\begin{bmatrix}
\vect{x}_1^{(2)} \\
\vect{x}_2^{(2)}
\end{bmatrix}=\begin{bmatrix}
-\matr{R}^{(2)} \\
\matr{0}
\end{bmatrix}\vect{u}_{2}^{(1)}
\label{eq:solve14a}
\\
 \vect{y}_{2}^{(1)}=\begin{bmatrix}
-\matr{R}^{(2)}\\
\matr{0}
\end{bmatrix}^T\begin{bmatrix}
\vect{x}_1^{(2)} \\
\vect{x}_2^{(2)}
\end{bmatrix}
\label{eq:solve14b}
\end{gather}
\end{subequations}
where
\begin{align}
\matr{C}_{22}^{(2)}&=\begin{bmatrix}
\matr{0} & \matr{I}_{m-2p}
\end{bmatrix}(\matr{Q}^{(2)})^T\matr{K}^{-1}\matr{C}_{22}\matr{K}^{-T}\matr{Q}^{(2)}\begin{bmatrix}
\matr{0}\\
\matr{I}_{m-2p}
\end{bmatrix} \label{eq:C222}
\\
\matr{C}_{21}^{(2)}&=\begin{bmatrix}
\matr{0} & \matr{I}_{m-2p}
\end{bmatrix}(\matr{Q}^{(2)})^T\matr{K}^{-1}\matr{C}_{22}\matr{K}^{-T}\matr{Q}^{(2)}\begin{bmatrix}
\matr{I}_{p} \\
\matr{0} 
\end{bmatrix}\\
\matr{C}_{11}^{(2)}&=\begin{bmatrix}
\matr{I}_{p} & \matr{0}
\end{bmatrix}(\matr{Q}^{(2)})^T\matr{K}^{-1}\matr{C}_{22}\matr{K}^{-T}\matr{Q}^{(2)}\begin{bmatrix}
\matr{I}_{p} \\
\matr{0} 
\end{bmatrix}
\end{align}
System~\eqref{eq:solve14a}-\eqref{eq:solve14b} is now in the form~\eqref{eq:solve8a}-\eqref{eq:solve8b}, and the reduction process used in iteration $1$ can be applied again. System~\eqref{eq:solve14a}-\eqref{eq:solve14b} can be seen as the cascade of a first system $\Sigma_{1}^{(2)}$ of order $p$
\begin{equation}\label{eq:solve16a}
\Sigma_{1}^{(2)}:\left\{
\begin{aligned}
\matr{I}_{p}\vect{x}_{1}^{(2)} + s\matr{C}_{11}^{(2)}\vect{x}_{1}^{(2)}&=\vect{u}_{1}^{(2)} - \matr{R}^{(2)}\vect{u}_{2}^{(1)}\\
\vect{y}_{2}^{(1)}&= -(\matr{R}^{(2)})^T\vect{x}_{1}^{(2)}
\end{aligned}
\right.
\end{equation}
and a second system $\Sigma_{2}^{(2)}$  of order $m-2p$
\begin{equation}\label{eq:solve16}
\Sigma_{2}^{(2)}:\left\{
\begin{aligned}
\matr{I}_{m-2p} \vect{x}_{2}^{(2)} + s\matr{C}_{22}^{(2)}\vect{x}_{2}^{(2)}&=-\matr{C}_{21}^{(2)}\vect{u}_{2}^{(2)}\\
\vect{y}_{2}^{(2)}&=-(\matr{C}_{21}^{(2)})^T\vect{x}_{2}^{(2)}
\end{aligned}
\right.
\end{equation}
The two systems are  only dynamically coupled, through equations $\vect{u}_{1}^{(2)}=s\vect{y}_{2}^{(2)}$ and $\vect{u}_{2}^{(2)}=s\vect{x}_{1}^{(2)}$. Overall, the original system~\eqref{eq:solve1} is now decomposed into three blocks, all coupled dynamically, as shown in Fig.~\ref{fig:interaction2}. If we retain the first two blocks, and neglect $\Sigma_{2}^{(2)}$, we obtained a ROM of order $2p$
\begin{subequations}\label{eq:solve17}
\begin{gather}
\left(
\begin{bmatrix}
\matr{G}_{11}^{(1)} & \matr{0}\\
\matr{0}& \matr{I}_{p}
\end{bmatrix} + s\begin{bmatrix}
\matr{C}_{11}^{(1)} & \ast\\
\matr{R}^{(2)} & \matr{C}_{11}^{(2)}
\end{bmatrix} \right)\begin{bmatrix}
\vect{x}_{1}\\
\vect{x}_{1}^{(2)}
\end{bmatrix}=\begin{bmatrix}
\matr{B}_{1}\\
\matr{0}
\end{bmatrix}\vect{u}\\
\hat{\vect{y}} = \begin{bmatrix}
\matr{B}_{1}^T &
\matr{0}
\end{bmatrix}
\begin{bmatrix}
\vect{x}_{1}\\
\vect{x}_{1}^{(2)}
\end{bmatrix}
\end{gather}
\end{subequations}
As shown in the Appendix, this model matches the first $4$ moments of the original system.

\subsubsection{Matching More Than Four Moments}

\begin{figure*}[!t]
\normalsize
\begin{align}\label{eq:solve19}
\left[\begin{IEEEeqnarraybox*}[][c]{,c/c/c/c,}
\matr{G}_{11}^{(1)} & & & \\
 &\matr{I}_{p}& & \\
  & &\ddots& \\
  & & &\matr{I}_{p}%
\end{IEEEeqnarraybox*}\right] \left[\begin{IEEEeqnarraybox}[][c]{,c,}
\vect{x}_{1} \\
\vect{x}_{1}^{(2)}\\
\vdots \\
\vect{x}_{1}^{(q)}
\end{IEEEeqnarraybox}\right] +
s\left[\begin{IEEEeqnarraybox*}[][c]{,c/c/c/c,}
\matr{C}_{11}^{(1)} &\ast &  & \\
 -\matr{R}^{(2)}&\matr{C}_{11}^{(2)}& \ddots & \\
  & \ddots & \ddots & \ast \\
  & & -\matr{R}^{(q)} & \matr{C}_{11}^{(q)}%
\end{IEEEeqnarraybox*}\right] \left[\begin{IEEEeqnarraybox}[][c]{,c,}
\vect{x}_{1} \\
\vect{x}_{1}^{(2)}\\
\vdots \\
\vect{x}_{1}^{(q)}
\end{IEEEeqnarraybox}\right]&=\left[\begin{IEEEeqnarraybox}[][c]{,c,}
\matr{B}_{1} \\
 \matr{0}\\
 \vdots\\
 \matr{0}
\end{IEEEeqnarraybox}\right]\vect{u}
\end{align}
\vspace*{4pt}
\hrulefill
\end{figure*}

Additional moments can be matched by iterating the proposed process, and further decompose subsystem $\Sigma_{2}^{(2)}$ in Fig.~\ref{fig:interaction2}. This goal can be achieved by computing, at each iteration $j \ge 3$, the QR decomposition
\begin{equation}
(\matr{Q}^{(j)})^T\matr{C}_{21}^{(j-1)}= \begin{bmatrix}
\matr{R}^{(j)}\\
\matr{0}
\end{bmatrix}
\end{equation}
of the input-to-state matrix of the innermost system (at iteration $j=3$, matrix $\matr{C}_{21}^{(2)}$ in~\eqref{eq:solve16}). The QR decomposition is obtained with a series of Householder reflectors that form the congruence matrix $\matr{Q}^{(j)}$. The obtained system will have the same structure as~\eqref{eq:solve14a}-\eqref{eq:solve14b}, and can be seen as the cascade of two blocks. The first system $\Sigma_{1}^{(j)}$, of size $p$, will add two matched moments to the ROM computed up to that point. The second system will be further decomposed if $j<q$. Otherwise, at the last iteration, it will be discarded. After $q$ iterations, the obtained ROM will have order $pq$, and will be in the form shown in equation~\eqref{eq:solve19} at the top of the next page. In the Appendix, we prove that the obtained model matches $2q$ moments of the original network. The proposed technique therefore leads to a ROM of the same size and accuracy as PRIMA, but in a more efficient way, which avoids the explicit construction of a huge and dense projection matrix. In comparison to SIP~\cite{ye2008sparse}, that can match only two moments per frequency point, the proposed method can match an arbitrary number of moments, and does not suffer from the singularity issues of multipoint SIP~\cite{ye2008sparse}. The use of PRIMA to match additional moments, advocated in SparseRC~\cite{ionuctiu2011sparserc}, is also avoided.

Another key advantage of the proposed method is the block-diagonal structure of~\eqref{eq:solve19}. Unlike PRIMA, that generates dense models, the proposed method naturally leads to a sparse representation. This reduces the memory footprint of the ROMs, and accelerates subsequent simulations, as we shall see in Sec.~\ref{NumericalResults}. Although PRIMA models can be sparsified with an eigenvalue decomposition, this operation costs extra CPU cycles. The obtained models are stable and passive by construction, since only congruence transformations like~\eqref{eq:prima1} have been used to generate the ROM matrices. The positive-definitive nature of $\matr{G}$ and $\matr{C}$ in~\eqref{eq:solve1} is thus preserved, which implies passivity and guarantees stable transient simulations~\cite{triverio2007stability}. We also note that TurboMOR preserved the matrix $\matr{B}_1$ in~\eqref{eq:solve7a} and~\eqref{eq:solve7b} that maps input ports to state equations. As discussed in~\cite{yang2007rlcsyn}, this property facilitates the connection of the ROM to the surrounding components. Finally, the obtained ROM can be converted into an RC equivalent circuit using the procedure in~\cite{yang2007rlcsyn}, for seamless integration into existing tools for electronic design automation.

\subsection{Practical Implementation}

We now discuss how TurboMOR can be implemented for maximum efficiency in terms of CPU time and memory consumption. The Cholesky decomposition of $\matr{G}_{22}$ can be obtained using efficient routines for the factorization of sparse, positive-definitive matrices, such as the supernodal method~\cite{suiteTim} available in MATLAB's \verb+chol+ routine. 
The QR decomposition in~\eqref{eq:qr} is computed with the Householder method. In our MATLAB implementation of TurboMOR, we used a direct call to the compiled LAPACK routine DGEQRF~\cite{netLAPACK}, which returns the orthogonal 
 $\matr{Q}^{(j)}$ matrix in factored form~\cite{golub2012matrix}. Such matrix is never computed explicitly, but kept in factored form. The LAPACK's routine DORMQR~\cite{netLAPACK} can be used to compute products involving $\matr{Q}^{(j)}$ directly from its factorization. Being large and dense, matrix $\matr{C}_{22}^{(j)}$ is also never computed explicitly. Its factored form is always used, which his given by~\eqref{eq:C222} for $j=2$ and by
\begin{equation}\label{eq:prac2}
 \matr{C}_{22}^{(j)}=\left[\begin{IEEEeqnarraybox}[][c]{,c/c,}
\matr{0} &\hspace{2mm} \matr{I}_{(m-jp)}
\end{IEEEeqnarraybox}\right](\matr{Q}^{(j)})^T\matr{C}_{22}^{(j-1)}\matr{Q}^{(j)}\left[\begin{IEEEeqnarraybox}[][c]{,c,}
\matr{0}\\
\matr{I}_{(m-jp)}
\end{IEEEeqnarraybox}\right]
\end{equation}
for $j >2$.

\subsection{On the Singularity of $\matr{G}_{22}$}
\label{sec:singularG}

Throughout the derivation of TurboMOR, we assumed the block $\matr{G}_{22}$ in~\eqref{eq:solve7a} to be strictly positive definite, hence invertible. When this is not the case, we adopt
the solution proposed in~\cite{ionuctiu2011sparserc} for SparseRC.
The rows and columns that make $\matr{G}_{22}$ singular are promoted into the first set of equations, and not eliminated. Since the number of such rows is typically very low, this does not significantly increase the size of the obtained ROMs.

\subsection{TurboMOR with partitioning}

Graph partitioning techniques can be integrated into TurboMOR to reduce very large networks, such as the power grid models that we will consider in Sec.~\ref{NumericalResults}. A possible partitioning strategy, used in~\cite{ionuctiu2011sparserc} and~\cite{miettinen2011partmor}, is to partition the given network into subnetworks that interact only through a limited set of nodes, called \emph{separator} nodes. An optimal partitioning can be found with the nested dissection algorithm \verb+nesdid+ from the SuiteSparse package~\cite{suiteTim}. Once the network nodes are reordered according to the partitions identified by  \verb+nesdis+, the matrices in~\eqref{eq:solve1} assume a bordered block diagonal form~\cite{zecevic1994balanced}.
 To illustrate this, consider a three-component partitioning of~\eqref{eq:solve1}
\begin{multline}\label{eq:partitioning1}
\left(
\begin{bmatrix}
\matr{G}_{1} & \matr{0} & \ast \\
\matr{0} & \matr{G}_{2} & \ast\\
\matr{G}_{31} & \matr{G}_{32} & \matr{G}_{3}
\end{bmatrix}
+ s\begin{bmatrix}
\matr{C}_{1} & \matr{0} & \ast\\
\matr{0} &\matr{C}_{2} & \ast\\
\matr{C}_{31} & \matr{C}_{32} & \matr{C}_{3}
\end{bmatrix} \right)
\\ \times \begin{bmatrix}
\vect{x}_1 \\
\vect{x}_2\\
\vect{x}_3
\end{bmatrix}=\begin{bmatrix}
\matr{B}_{1}  \\ \matr{B}_{2} \\ \matr{B}_{3}
\end{bmatrix} \vect{u}
\end{multline}
Blocks $\matr{G}_1, \matr{C}_1$ and $\matr{G}_2, \matr{C}_2$ correspond to two decoupled subsystems, that interact only through a set of separator nodes associated to $\matr{G}_3$, via coupling matrices $\matr{G}_{31}$, $\matr{C}_{31}$, $\matr{G}_{32}$, $\matr{C}_{32}$. Subsystems $1$ and $2$ can be reduced individually. The coupling matrices are then updated accordingly. For instance, for reducing subsystem $1$, we first form its nodal equations
\begin{equation}\label{eq:partitioning2}
\left(
\begin{bmatrix}
\matr{G}_{1} & \ast \\
\matr{G}_{31}  & \matr{G}_{3}
\end{bmatrix}
+ s\begin{bmatrix}
\matr{C}_{1}  & \ast\\
\matr{C}_{31} & \matr{C}_{3}
\end{bmatrix} \right)
\begin{bmatrix}
\vect{x}_1 \\
\vect{x}_3
\end{bmatrix} =\begin{bmatrix}
\matr{B}_{1} \\
\matr{B}_{3}
\end{bmatrix}\vect{u}
\end{equation}
and then reorder its nodes such that
\begin{itemize}
\item port nodes and separator nodes come first, and form the state vector $\vect{x}_1$ in~\eqref{eq:solve7a};
\item internal nodes come second, forming $\vect{x}_2$ in~\eqref{eq:solve7a}.
\end{itemize}

Then, we perform the reduction as in Sec.~\ref{sec:theory}. After all subsystems have been reduced, the obtained ROM will read
\begin{equation} \label{eq:partitioning4}
\left(
\begin{bmatrix}
\hat{\matr{G}}_{1} & \matr{0} & \ast \\
\matr{0} & \hat{\matr{G}}_{2} & \ast\\
\hat{\matr{G}}_{31} & \hat{\matr{G}}_{32} & \tilde{\matr{G}}_{3}
\end{bmatrix} \!\!
+ \!\! s\begin{bmatrix}
\hat{\matr{C}}_{1} & \matr{0} & \ast\\
\matr{0} &\hat{\matr{C}}_{2} & \ast\\
\hat{\matr{C}}_{31} & \hat{\matr{C}}_{32} & \tilde{\matr{C}}_{3}
\end{bmatrix} \right) 
\!\!\!\!
\begin{bmatrix}
\hat{\vect{x}}_1 \\
\hat{\vect{x}}_2\\
\vect{x}_3
\end{bmatrix} \!\! = \!\! \begin{bmatrix}
\hat{\matr{B}}_{1} \\ \hat{\matr{B}}_{2} \\ \matr{B}_{3}
\end{bmatrix} \vect{u}
\end{equation}
As numerical results will show, partitioning reduces the overall cost of the reduction, since TurboMOR is applied to subsystems of smaller size. Additionally, it reduces the number of fill-ins in the ROM, since the zero blocks in~\eqref{eq:partitioning1} are maintained in~\eqref{eq:partitioning4}.

\section{Numerical Results}\label{NumericalResults}

The proposed TurboMOR algorithm has been implemented in MATLAB, with direct calls to compiled LAPACK libraries for a few key operations, namely the QR decomposition of~\eqref{eq:qr}, and the computation of the products with the Householder matrices $\matr{Q}^{(j)}$. In this section, we compare the performance of TurboMOR against PRIMA~\cite{odabasioglu1997prima} and SparseRC~\cite{ionuctiu2011sparserc}. Computations were performed on a $3.40$~GHz Intel~i7 CPU, with 16~GB of memory and MATLAB R2013b. 

\subsection{Reduction Time}

\begin{table*}[!t]
\renewcommand{\arraystretch}{1.3}
\caption{Reduction time for the different methods on various test networks. All times are in seconds.}
\label{tab:table_example1}
\centering
\begin{tabular}{|m{1.9cm}<{\centering}|m{0.2cm}<{\centering}|m{1.0cm}<{\centering}|m{1.0cm}<{\centering}|m{1.6cm}<{\centering}|m{1.0cm}<{\centering}|m{1.6cm}<{\centering}|m{1.0cm}<{\centering}|m{1.6cm}<{\centering}|}
\hline
\multirow{2}{*}{Examples}&\multirow{2}{*}{$q$}&\multirow{2}{*}{PRIMA}&\multicolumn{2}{c|}{SparseRC}&\multicolumn{2}{c|}{TurboMOR}&\multicolumn{2}{c|}{TurboMOR with partitioning}\\
\cline{4-9}
& &cpu time &cpu time&Speedup w.r.t PRIMA&cpu time&Speedup w.r.t PRIMA&cpu time&Speedup w.r.t PRIMA\\
\hline \hline
\multirow{3}{1.9cm}{\centering $1$. On-chip bus $p=256$ $m=38,528$}&$1$&$0.94$&$0.40$&$2.35 \times$&$0.28$&$3.36 \times$&$0.37$&$2.54 \times$\\
& $2$&$2.84$&$1.83$&$1.55 \times$&$1.48$&$1.92 \times$&$1.58$&$1.80 \times$\\
&$3$&$4.78$&$3.63$&$1.32 \times$&$3.00$&$1.59 \times$&$2.94$&$1.63 \times$\\
\hline
\multirow{3}{2.0cm}{\centering $2$. ibmpg1t (RC) $p=200$ $m=25,195$}&$1$&$0.41$&$0.16$&$2.56 \times$&$0.19$&$2.16 \times$&$0.18$&$2.28 \times$\\
& $2$&$1.10$&$0.51$&$2.16 \times$&$0.66$&$1.67 \times$&$0.45$&$2.44 \times$\\
&$3$&$1.98$&$0.94$&$2.11 \times$&$1.38$&$1.43 \times$&$0.85$&$2.33 \times$\\
\hline
\multirow{3}{2.0cm}{\centering $3$. ibmpg2t (RC) $p=800$ $m=163,697$}&$1$&$22.00$&$6.24$&$3.53 \times$&$10.84$&$2.03 \times$&$6.28$&$3.50 \times$\\
& $2$&$65.55$&$20.89$&$3.14 \times$&$37.82$&$1.73 \times$&$18.89$&$3.47 \times$\\
&$3$&$118.64$&$39.48$&$3.01 \times$&$78.38$&$1.51 \times$&$35.28$&$3.36 \times$\\
\hline
\multirow{3}{2.0cm}{\centering $4$. ibmpg2t (RC) $p=1200$ $m=163,697$}&$1$&$35.51$&$9.53$&$3.73 \times$&$16.64$&$2.13 \times$&$9.52$&$3.73 \times$\\
& $2$&$109.41$&$32.63$&$3.35 \times$&$60.60$&$1.81 \times$&$29.19$&$3.75 \times$\\
&$3$&$224.06$&$64.41$&$3.48 \times$&$132.55$&$1.69 \times$&$56.58$&$3.96 \times$\\
\hline
\multirow{3}{2.0cm}{\centering $5$. ibmpg2t (RC) $p=1500$ $m=163,697$}&$1$&$49.50$&$11.66$&$4.25 \times$&$21.29$&$2.33 \times$&$11.76$&$4.21 \times$\\
& $2$&$152.81$&$43.06$&$3.55 \times$&$83.66$&$1.83 \times$&$37.58$&$4.07 \times$\\
&$3$&$729.92$&$83.76$&$8.71 \times$&$186.95$&$3.90 \times$&$72.32$&$10.09 \times$\\
\hline
\multirow{3}{2.0cm}{\centering $6$. ibmpg2t (RC) $p=2000$ $m=163,697$}&$1$&$73.17$&$16.29$&$4.49 \times$&$31.23$&$2.34 \times$&$16.29$&$4.49 \times$\\
& $2$&$340.18$&$62.12$&$5.48 \times$&$228.32$&$1.49 \times$&$54.76$&$6.21 \times$\\
&$3$&$9807.11$&$122.12$&$80.31 \times$&$1051.72$&$9.32 \times$&$115.71$&$84.76 \times$\\
\hline
\end{tabular}
\end{table*}
Table~\ref{tab:table_example1} shows the time needed by the different methods to reduce various test networks. Example $1$ is an on-chip bus consisting of $128$ signal lines.
The bus was modelled with lumped RC segments, and has the characteristics of a global interconnect in the $65$nm technology node~\cite{onChip}.
Examples $2$ - $6$ are power grid benchmarks obtained from~\cite{ibmpg}. The original benchmarks include some inductors, which were neglected. A variable number of input current sources has been considered to investigate the scalability of the MOR methods with respect to port count.

We first compare the proposed method \textit{without} partitioning against PRIMA, in order to assess its intrinsic efficiency in matching moments. For each test case, reduced order models have been generated to match 2, 4, and 6 moments. From the results in Table~\ref{tab:table_example1}, we observe that TurboMOR is consistently faster than PRIMA, up to 9.32~times. Savings are particularly high when order and port count are high, as in example~6. While PRIMA takes 2~hours and 43~minutes (9807~s) to match 6~moments, TurboMOR achieves the same result in only 17.5~minutes (1051~s).
This speed-up is due to the fact that TurboMOR achieves moment matching without computing and storing a large projection matrix as PRIMA does.

Then, we compare TurboMOR \textit{with} partitioning against the recently-proposed SparseRC method~\cite{ionuctiu2011sparserc}. From Table \ref{tab:table_example1}, we observe that partitioning improves reduction time substantially, especially for large networks (examples 3, 4, 5 and~6). 
Comparing the proposed method and SparseRC, we see that for two moments matched ($q=1$), both methods have almost the same reduction time. This is expected since, in this case,  the methods perform the same operations. However, when additional moments are matched ($q=2$ and $q=3$), the proposed method is always faster than SparseRC, which employs PRIMA to match additional moments, losing some efficiency. This result shows how, with the Householder transformations proposed in Sec.~\ref{sec:theory}, additional moments can be efficiently matched.

\subsection{Accuracy of the Reduced Models}
\begin{figure}[!t]
\centering
\includegraphics[width=0.95\columnwidth]{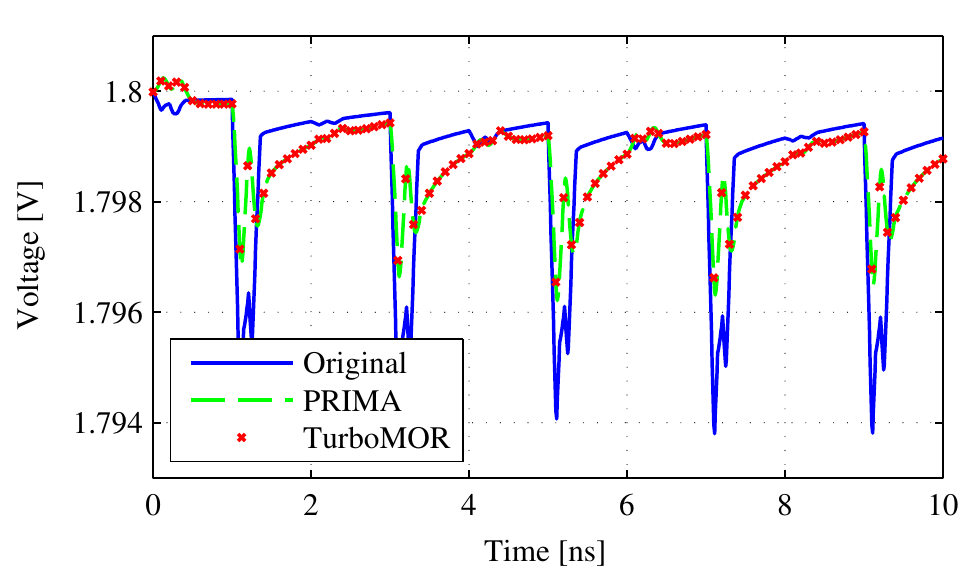}
\caption{Transient response of the original system and the reduced models obtained with TurboMOR and PRIMA. The reduced models match two moments ($q=1$).}
\label{fig:fig_sim1}
\end{figure}

\begin{figure}[!t]
\centering
\includegraphics[width=0.95\columnwidth]{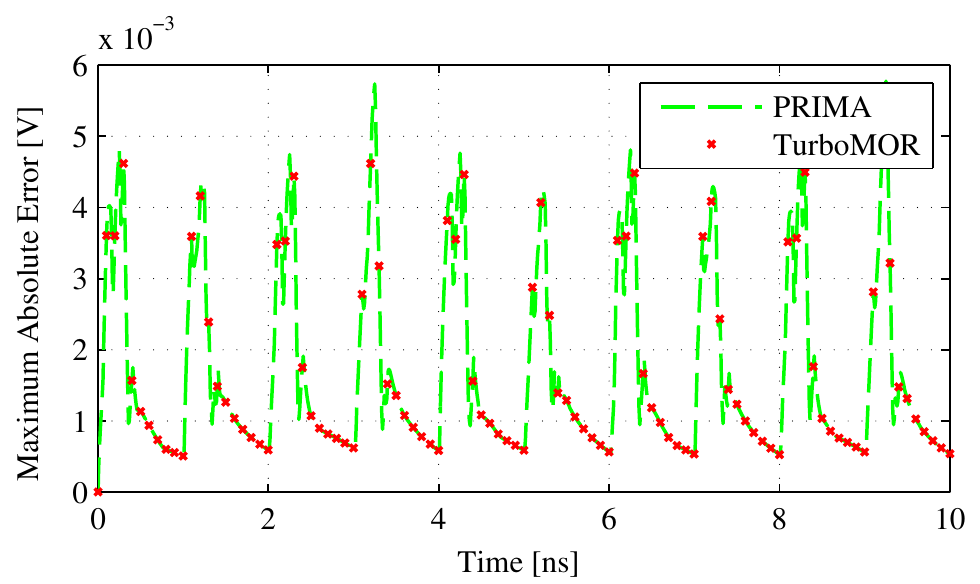}
\caption{Error between the response of the original system and the response of the reduced models computed with PRIMA and the proposed method. The reduced models match two moments ($q=1$).}
\label{fig:fig_sim2}
\end{figure}

\begin{figure}[!t]
\centering
\includegraphics[width=0.95\columnwidth]{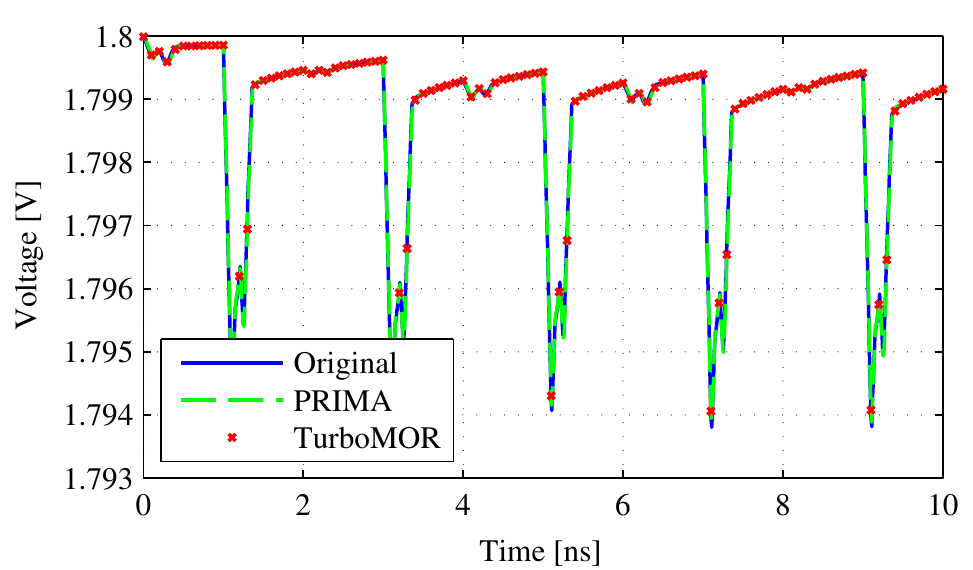}
\caption{As in Fig.~\ref{fig:fig_sim1}, but for four moments matched ($q=4$).}
\label{fig:fig_sim3}
\end{figure}

\begin{figure}[!t]
\centering
\includegraphics[width=0.95\columnwidth]{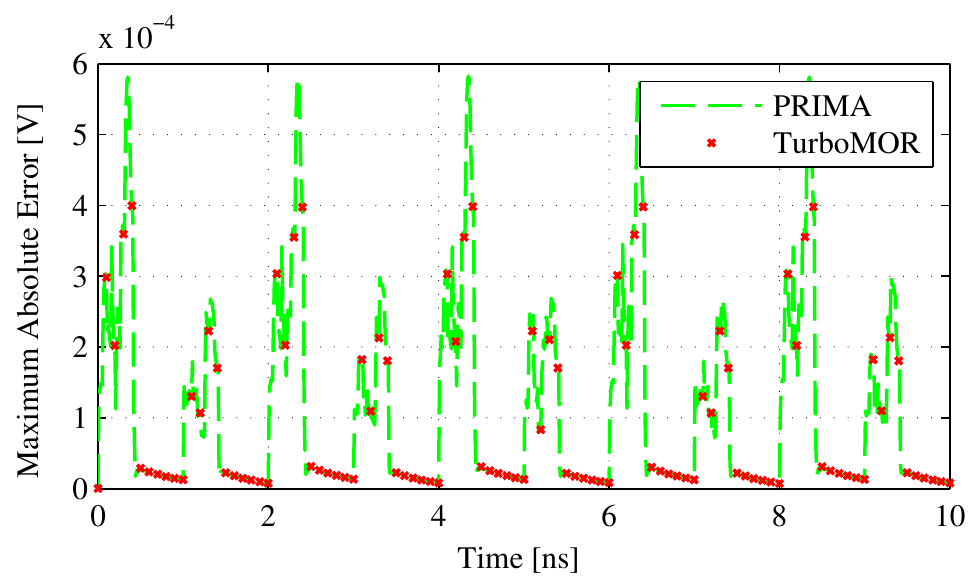}
\caption{As in Fig.~\ref{fig:fig_sim2}, but for four moments matched ($q=4$).}
\label{fig:fig_sim4}
\end{figure}

In this section we demonstrate that, from an accuracy standpoint, TurboMOR is equivalent to PRIMA. 
 For this purpose, we consider the power grid ``ibmpg1t'' from~\cite{nassif2008power}, which corresponds to example~2 in Table~\ref{tab:table_example1}. A transient simulation is performed to calculate the voltage at one of the supply ports of the power grid, when switching currents are drawn by the different blocks of the integrated circuit. Fig.~\ref{fig:fig_sim1} shows the time response obtained with the original system and the reduced models from TurboMOR and PRIMA, for the case of two moments matched ($q=1$). Both methods provide similar results. This confirms that the proposed method is as accurate as PRIMA, but more efficient. 
 
In Fig.~\ref{fig:fig_sim2}, the maximum error for the two ROMs is depicted. Figures show that a ROM with only two moments matched is not suitable for an accurate assessment of the voltage drop across the power grid. Indeed, the ROMs underestimate the voltage drop, by as much as 5~mV.  
In Fig.~\ref{fig:fig_sim3}, we show the transient results obtained with PRIMA and TurboMOR models that match four moments ($q=2$). Now, both models lead to a very accurate prediction of the original system response. The worst case transient error is indeed below 1~mV, as shown by Fig.~\ref{fig:fig_sim4}. This example shows that matching only two moments as in SIP~\cite{ye2008sparse} is not accurate enough for some applications. TurboMOR can instead match an arbitrary number of moments, and meet any accuracy requirement set by the user.

\subsection{Efficiency of the Reduced Models}

\begin{table*}[!t]
\renewcommand{\arraystretch}{1.3}
\caption{Simulation time for the ROMs obtained with the different methods. All times in seconds.}
\label{tab:table_example2}
\centering
\begin{tabular}{|m{2.0cm}<{\centering}|m{1.0cm}<{\centering}|m{0.2cm}<{\centering}|m{1.0cm}<{\centering}|m{1.0cm}<{\centering}|m{1.0cm}<{\centering}|m{1.0cm}<{\centering}|m{1.0cm}<{\centering}|m{1.6cm}<{\centering}|m{1.0cm}<{\centering}|m{1.2cm}<{\centering}|}
\hline
\multirow{2}{*}{Examples}&\multirow{2}{*}{Original}&\multirow{2}{*}{$q$}&\multicolumn{2}{c|}{PRIMA}&\multicolumn{2}{c|}{SparseRC}&\multicolumn{2}{c|}{TurboMOR}&\multicolumn{2}{c|}{TurboMOR with partitioning}\\
\cline{4-11}
&Sim.Time& &Sim.Time&Speedup&Sim.Time&Speedup&Sim.Time&Speedup&Sim.Time&Speedup\\
\hline \hline
\multirow{3}{2.0cm}{\centering $1$. On-chip bus $p=256$ $m=38,528$}&\multirow{2}{1.0cm}{$3.00$}&$1$&$0.07$&$42.86 \times$&$0.06$&$50.00 \times$&$0.07$&$42.86 \times$&$0.06$&$50.00 \times$\\
& &$2$&$0.76$&$3.95 \times$&$0.80$&$3.75 \times$&$0.24$&$12.50 \times$&$0.60$&$5.00 \times$\\
& &$3$&$3.23$&$0.93 \times$&$2.38$&$1.26 \times$&$0.54$&$5.56 \times$&$1.48$&$2.03 \times$\\
\hline
\multirow{3}{2.0cm}{\centering $2$. ibmpg1t (RC) $p=200$ $m=25,195$}&\multirow{2}{1.0cm}{$2.26$}&$1$&$0.23$&$9.83 \times$&$0.14$&$16.14\times$&$0.13$&$17.38 \times$&$0.14$&$16.14 \times$\\
& &$2$&$0.76$&$2.97 \times$&$0.32$&$7.06 \times$&$0.33$&$6.85 \times$&$0.27$&$8.37 \times$\\
& &$3$&$1.92$&$1.18 \times$&$1.19$&$1.90 \times$&$0.61$&$3.70 \times$&$0.58$&$3.90 \times$\\
\hline
\multirow{3}{2.0cm}{\centering $3$. ibmpg2t (RC) $p=800$ $m=163,697$}&\multirow{2}{1.0cm}{$29.84$}&$1$&$4.32$&$6.91 \times$&$1.58$&$18.89 \times$&$1.60$&$18.65 \times$&$1.61$&$18.53 \times$\\
& &$2$&$12.98$&$2.30 \times$&$5.94$&$5.02 \times$&$8.11$&$3.68 \times$&$4.78$&$6.24 \times$\\
& &$3$&$27.28$&$1.09 \times$&$12.68$&$2.35 \times$&$13.63$&$2.19 \times$&$7.55$&$3.95 \times$\\
\hline
\multirow{3}{2.0cm}{\centering $4$. ibmpg2t (RC) $p=1200$ $m=163,697$}&\multirow{2}{1.0cm}{$30.24$}&$1$&$8.58$&$3.52 \times$&$3.56$&$8.49 \times$&$3.67$&$8.24 \times$&$3.56$&$8.49 \times$\\
& &$2$&$28.75$&$1.05 \times$&$12.98$&$2.33 \times$&$20.30$&$1.49 \times$&$10.10$&$2.99 \times$\\
& &$3$&$62.51$&$0.48 \times$&$28.40$&$1.06 \times$&$34.98$&$0.86 \times$&$16.21$&$1.87 \times$\\
\hline
\multirow{3}{2.0cm}{\centering $5$. ibmpg2t (RC) $p=1500$ $m=163,697$}&\multirow{2}{1.0cm}{$30.69$}&$1$&$13.31$&$2.31 \times$&$5.55$&$5.53 \times$&$5.98$&$5.13 \times$&$5.60$&$5.48 \times$\\
& &$2$&$45.32$&$0.68 \times$&$20.27$&$1.51 \times$&$35.23$&$0.87 \times$&$15.55$&$1.97 \times$\\
& &$3$&$104.12$&$0.29 \times$&$42.93$&$0.71 \times$&$60.40$&$0.51 \times$&$24.21$&$1.27 \times$\\
\hline
\multirow{3}{2.0cm}{\centering $6$. ibmpg2t (RC) $p=2000$ $m=163,697$}&\multirow{2}{1.0cm}{$30.80$}&$1$&$23.08$&$1.33 \times$&$9.45$&$3.26 \times$&$10.89$&$2.83 \times$&$9.58$&$3.22 \times$\\
& &$2$&$81.00$&$0.38 \times$&$34.84$&$0.88 \times$&$73.17$&$0.42 \times$&$26.80$&$1.15 \times$\\
& &$3$&$173.67$&$0.18 \times$&$77.28$&$0.40 \times$&$121.54$&$0.25 \times$&$43.18$&$0.71 \times$\\
\hline
\end{tabular}
\end{table*}

We now evaluate the efficiency of the ROMs generated by the proposed method, PRIMA, and SparseRC. In Table~\ref{tab:table_example2}, the simulation time for the original network and the various ROMs is reported.

\textit{Without} partitioning, TurboMOR produces ROMs that are consistently faster than PRIMA models. This is attributed to the block diagonal structure of the reduced models, which reduces the cost of the LU~factorizations used to perform subsequent transient simulations. TurboMOR models are faster by up to five times.

Comparing now the simulation times for the methods \textit{with} partitioning (SparseRC and TurboMOR with partitioning), we  observe that when two moments are matched ($q=1$), the simulation times are essentially the same, which is expected since both methods adopt the same reduction strategy. However, when additional moments are matched, TurboMOR delivers models that are always faster than those from SparseRC, because of higher sparsity. SparseRC uses PRIMA to match additional moments, which introduces some large and dense blocks in the ROM.

\subsection{Scalability}

\begin{figure}[!t]
\centering
\includegraphics[width=0.95\columnwidth]{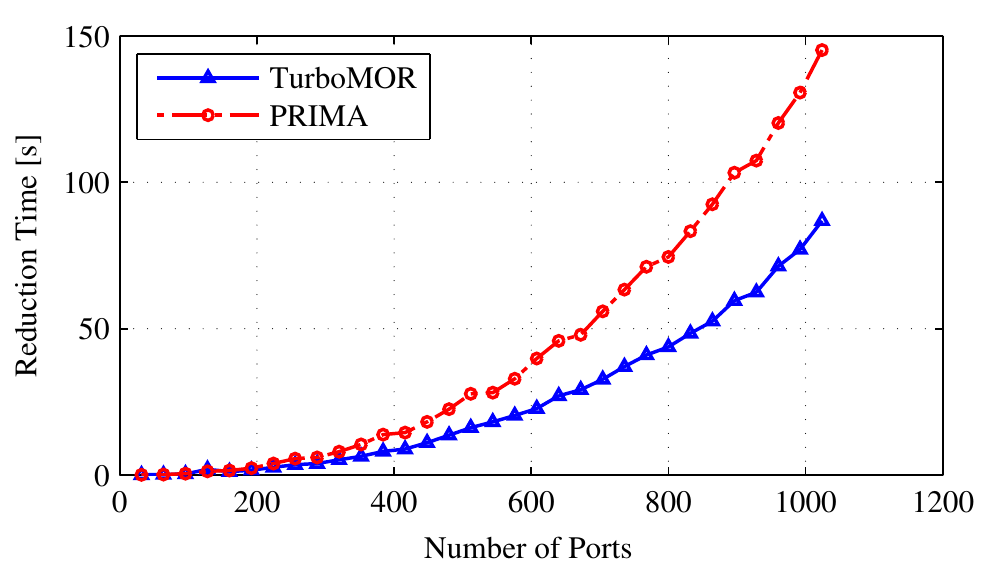}
\caption{Reduction time for PRIMA and TurboMOR without partitioning vs number of ports. Both methods match six moments ($q=3$).}
\label{fig:fig_sim5}
\includegraphics[width=0.95\columnwidth]{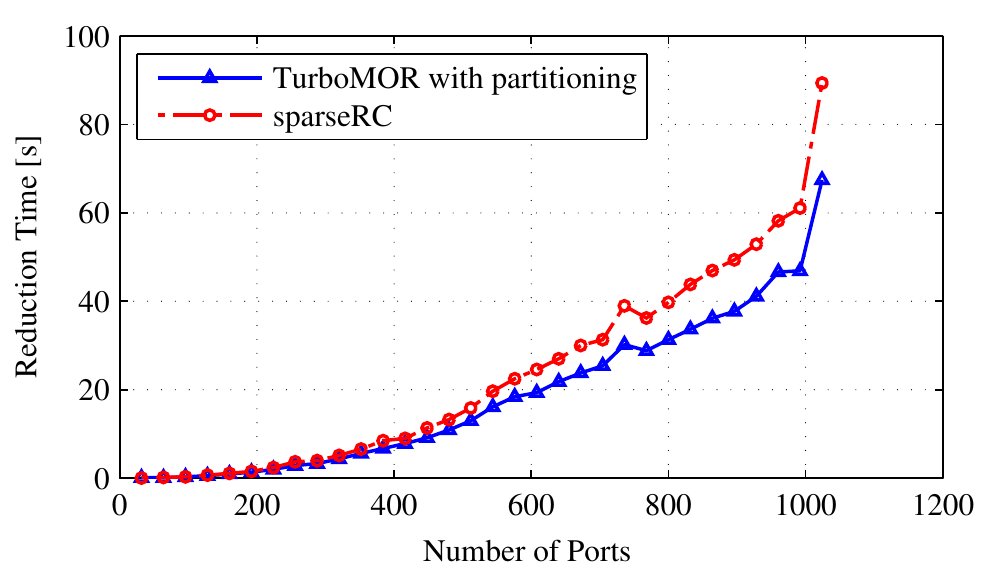}
\caption{Reduction time for SparseRC and TurboMOR with partitioning vs number of ports. Both methods match six moments ($q=3$).}
\label{fig:fig_sim6}
\end{figure}

Finally, we investigate the scalability of TurboMOR and existing methods with respect to network order and number of ports. Tests are performed on the first example (on-chip bus) for the case of six moments matched.

\subsubsection{Varying Number of Ports, Constant Node-to-Port Ratio}

In the first test, we vary the number of signal lines and, consequently, ports. Since bus length is kept constant, the network order increases linearly with the number of ports. The node-to-port ratio remains constant at $150.5$.

Fig.~\ref{fig:fig_sim5} depicts the reduction time for TurboMOR (without partitioning) and PRIMA versus the number of ports. We observe that TurboMOR scales better than PRIMA, and time savings grow as port count increases. In Fig.~\ref{fig:fig_sim6}, the analysis is repeated for the proposed method with partitioning and SparseRC. Also in this case, TurboMOR scales better than existing methods.

\subsubsection{Varying Node-to-Port Ratio, Constant Number of Ports}
\begin{figure}[!t]
\centering
\includegraphics[width=0.95\columnwidth]{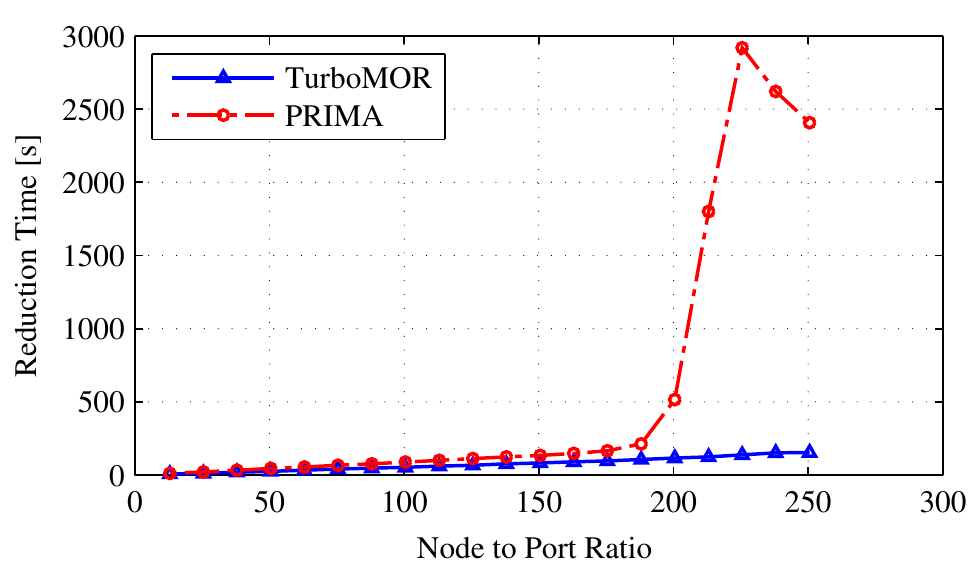}
\caption{Reduction time for PRIMA and proposed method \textit{without} partitioning, as a function of the ratio of network order and number of ports.}
\label{fig:fig_sim7}
\includegraphics[width=0.95\columnwidth]{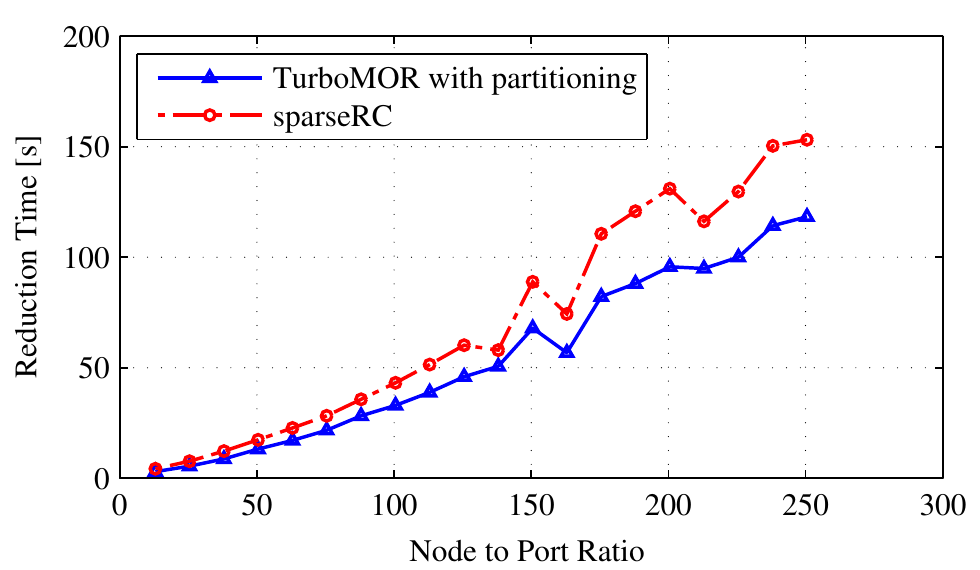}
\caption{As in Fig.~\ref{fig:fig_sim7}, but for SparseRC and proposed method \emph{with} partitioning.}
\label{fig:fig_sim8}
\end{figure}

In the second test, we keep the number of ports constant to 1024, which corresponds to 512 lines. We increase the number of nodes and, consequently,  order by making the bus longer. 

Fig.~\ref{fig:fig_sim7} shows the reduction time for the two methods without partitioning (proposed and PRIMA) as a function of the number of nodes. Beyond a certain point, the reduction time for PRIMA increases dramatically, because the projection matrix becomes larger than the 16~GB of memory available on the machine. PRIMA starts resorting to slow swap memory, and becomes very inefficient. With TurboMOR, large projection matrices are avoided. The matrices used to perform the congruence transformations are either sparse (Cholesky factor $\matr{K}$) or stored in efficient factored form (Householder reflectors in $\matr{Q}^{(i)}$). This results in lower memory consumption, and allows TurboMOR to achieve high scalability even for very large port counts. In Fig.~\ref{fig:fig_sim8}, the analysis is repeated for TurboMOR \emph{with} partitioning and SparseRC. The figure confirms the efficiency of the proposed models, which are faster than those generated by SparseRC especially for large systems with many ports.

\section{Conclusion}
\label{Conclusion}

We introduced TurboMOR, a new model order reduction method for large RC networks with many ports. TurboMOR achieves moment matching via efficient Householder transformations, sparse matrix factorizations, and graph partitioning techniques. Differently from popular methods such as PRIMA, no large and dense projection matrices need to be computed nor stored. This feature makes TurboMOR more efficient than existing methods in terms of both CPU time and memory consumption.
A key novelty of the proposed method is the sparse and block-diagonal structure of the generated models, which makes them faster at run-time. Based on this structure, we provide a nice interpretation of moment matching in terms of system theory. TurboMOR models are passive by construction, and can be cast into an  equivalent RC circuit, for seamless integration into electronic design automation tools. Numerical results demonstrate the superior performance of TurboMOR in reducing large passive networks with many ports, that arise more and more frequently in practice.

\appendices

\section{Proof of moment matching}\label{proof}

We prove that the reduced model~\eqref{eq:solve19}, obtained after $q$ iterations of the proposed method, matches $2q$ moments. We assume $\matr{G}$ invertible, since otherwise moments~\eqref{eq:M} are 
not defined. If $\matr{G}$ is singular, the proposed method will still work, but one cannot speak of moment matching.

The starting point of the proof is realization~\eqref{eq:solve8a}-\eqref{eq:solve8b}, which is obtained from the original system~\eqref{eq:solve7a}-\eqref{eq:solve7b} by means of congruence transformation~\eqref{eq:congruence}. Since~\eqref{eq:congruence} is invertible by construction, the transformation does not change the transfer function nor the system moments. 

The key argument of the proposed proof is the derivation of the relation between the moments $\matr{M}_k$ of the original system~\eqref{eq:solve8a}-\eqref{eq:solve8b} and the moments of the inner subsystem~\eqref{eq:solve11} extracted by TurboMOR after one iteration.
The transfer function of the original system~\eqref{eq:solve8a}-\eqref{eq:solve8b} can be written as~\cite{kerns1997stable,ionuctiu2011sparserc}
\begin{multline} \label{eq:Hdecomp}
	\matr{H}(s) = \matr{B}_1^T 
	\bigg[ \matr{G}_{11}^{(1)} + s \matr{C}_{11}^{(1)} 
	- s^2 \matr{H}_1(s) \bigg]^{-1} 
	\matr{B}_1
\end{multline}
where
\begin{equation} \label{eq:H2}
	\matr{H}_1(s) = \left( \matr{C}_{21}^{(1)} \right)^T 
	(\matr{G}_{22} + s \matr{C}_{22})^{-1} \matr{C}_{21}^{(1)}
\end{equation}
is the transfer function of the inner subsystem $\Sigma_2^{(1)}$.  
The moments of this subsystem are denoted with $\matr{N}_l$, so we have
\begin{equation} \label{eq:H2moments}
	\matr{H}_1 (s) = \sum_{l=0}^{+\infty} \matr{N}_l s^l
\end{equation}
After substituting~\eqref{eq:M} and~\eqref{eq:H2moments} into~\eqref{eq:Hdecomp}, we obtain
\begin{equation}
	\sum_{k=0}^{+\infty} \matr{M}_k s^k = \matr{B}_1^T 
	\bigg[ \matr{G}_{11}^{(1)} + s \matr{C}_{11}^{(1)} 
	-  \sum_{l=0}^{+\infty} \matr{N}_l s^{l+2} \bigg]^{-1} 
	\matr{B}_1
	\label{eq:Hdecomp2}
\end{equation}
For circuits, matrix $\matr{B}_1$ is typically a permutation of the identity matrix, and is thus invertible\footnote{If $\matr{B}_1$ is not full rank, a correlation between some inputs exists, which can be extracted before the reduction~\cite{feldmann2004model}, making the ROM smaller and leading to a full-rank $\matr{B}_1$.}. We can thus rewrite~\eqref{eq:Hdecomp2} as
\begin{equation}
	\bigg[ \matr{G}_{11}^{(1)} + s \matr{C}_{11}^{(1)} 
	-  \sum_{l=0}^{+\infty} \matr{N}_l s^{l+2} \bigg] \sum_{k=0}^{+\infty} \matr{B}_1^{-T} \matr{M}_k s^k =
	\matr{B}_1
	\label{eq:Hdecomp3}
\end{equation}
where superscript $^{-T}$ denotes the inverse of the transpose. After exchanging the two series, we have
\begin{multline}
	\sum_{k=0}^{+\infty} \bigg [ \matr{G}_{11}^{(1)} \matr{B}_1^{-T} \matr{M}_{k} s^k
	+ \matr{C}_{11}^{(1)} \matr{B}_1^{-T} \matr{M}_k s^{k+1} \\
	- \sum_{l=0}^{+\infty} \matr{N}_l \matr{B}_1^{-T} \matr{M}_k s^{k+l+2} \bigg] = \matr{B}_1 \label{eq:Hdecomp4}
\end{multline}
Both sides of~\eqref{eq:Hdecomp4} are polynomials in $s$ that, in order to be equal, must have the same coefficients. Imposing the equality between the coefficients of $s^0$ we obtain
\begin{equation}
	\matr{G}_{11}^{(1)} \matr{B}_1^{-T} \matr{M}_0 = \matr{B}_1 \Rightarrow \matr{M}_0 = \matr{B}_1^T \Big(  \matr{G}_{11}^{(1)} \Big)^{-1} \matr{B}_1
	\label{eq:M0}
\end{equation}
The inverse of $\matr{G}_{11}^{(1)}$ exists since we $\matr{G}$ is non-singular. By equating the coefficients of $s^1$, we have
\begin{equation}
	\matr{M}_{1}  =-\matr{B}_{1}^{T}(\matr{G}_{11}^{(1)})^{-1}\matr{C}_{11}^{(1)}(\matr{G}_{11}^{(1)})^{-1}\matr{B}_{1}
	\label{eq:M1}
\end{equation}
Equations~\eqref{eq:M0} and~\eqref{eq:M1} show that the first two moments of the original system just depend on the matrices $\matr{B}_{1}$, $\matr{G}_{11}^{(1)}$ and $\matr{C}_{11}^{(1)}$. Such matrices are preserved in reduced model~\eqref{eq:solve12}, which thus matches the first two moments of the original system.
By equating the coefficients of a generic power $s^r$ in~\eqref{eq:Hdecomp4} for $r \ge 2$, we obtain the recursive relation
\begin{multline}
	\matr{M}_r = - \matr{B}_1^T \Big( \matr{G}_{11}^{(1)} \Big)^{-1} \matr{C}_{11}^{(1)} \matr{B}_1^{-T} \matr{M}_{r-1} \\ + \matr{B}_1^T \Big( \matr{G}_{11}^{(1)} \Big)^{-1} \sum_{l=0}^{r-2} \matr{N}_l \matr{B}_1^{-T} \matr{M}_{r-l-2}
	\label{eq:Mr}
\end{multline}
Equation~\eqref{eq:Mr} shows that the moment $\matr{M}_r$ of order $r$ of the original system~\eqref{eq:solve8a}-\eqref{eq:solve8b} depends on:
\begin{enumerate}
\item the matrices $\matr{B}_{1}$, $\matr{G}_{11}^{(1)}$ and $\matr{C}_{11}^{(1)}$ of the outer subsystem~\eqref{eq:solve11a}, which are always preserved in the reduced model~\eqref{eq:solve19};
\item the moments $\matr{N}_l$ of the inner subsystem~\eqref{eq:solve11} up to order $r-2$.
\end{enumerate}
Therefore, if one replaces the \emph{nested} subsystem~\eqref{eq:solve11} with a reduced model that preserves its first $r-2$ moments, then the overall model will match $r$ moments of the original system.  By iterating this argument, it is straightforward to prove that ROM~\eqref{eq:solve19} matches $2q$ moments of the original system. 

The developed relation between the moments of the original system and the moments of its inner subsystem~\eqref{eq:solve11} plays a fundamental role in the proposed method. It allows us to match moments recursively, two at a time, by iterative application of the same transformation to subsystems of decreasing size. The proposed proof is also applicable to the ROMs obtained from other techniques such as SparseRC~\cite{ionuctiu2011sparserc}. The main differences between our proof and the one in~\cite{ionuctiu2011sparserc} are two. First, the proof in~\cite{ionuctiu2011sparserc} considers only the first two moments, while ours is general. Second, \cite{ionuctiu2011sparserc} proves moment matching for the moments of the network \emph{admittance}. Our proof is instead based on the original  \emph{impedance} representation of network~\eqref{eq:solve1}. Our contribution therefore establishes the equivalence, from a moment-matching perspective, of fast MOR methods (proposed, SparseRC) and PRIMA.

\ifCLASSOPTIONcaptionsoff
  \newpage
\fi



%


\bibliographystyle{IEEEtran}
\bibliography{IEEEabrv,mybib2}

\end{document}